# The Mid-Frequency Square Kilometre Array Phase Synchronisation System

S. W. Schediwy[1,2,*], D. R. Gozzard[1,2], C. Gravestock[1], S. Stobie[1], R. Whitaker[3], J. A. Malan[4], P. Boven[5] and K. Grainge[3]
[1]International Centre for Radio Astronomy Research, School of Physics, Mathematics & Computing, The University of Western Australia, Perth, WA 6009, Australia
[2]Department of Physics, School of Physics, Mathematics & Computing, The University of Western Australia, Perth, WA 6009, Australia
[3]Jodrell Bank Centre for Astrophysics, School of Physics & Astronomy, The University of Manchester, Manchester, M13 9PL, UK
[4]Square Kilometre Array South Africa, The South African Radio Astronomy Observatory, Pinelands 7405, South Africa
[5]Joint Institute for VLBI ERIC (JIVE), Dwingeloo, The Netherlands

## Abstract

This paper describes the technical details and practical implementation of the phase synchronisation system selected for use by the Mid-Frequency Square Kilometre Array (SKA). Over a four-year period, the system has been tested on metropolitan fibre-optic networks, on long-haul overhead fibre at the South African SKA site, and on existing telescopes in Australia to verify its functional performance. The tests have shown that the system exceeds the 1-second SKA coherence loss requirement by a factor of 2560, the 60-second coherence loss requirement by a factor of 239, and the 10-minute phase drift requirement by almost five orders-of-magnitude. The paper also reports on tests showing that the system can operate within specification over all the required operating conditions, including maximum fibre link distance, temperature range, temperature gradient, relative humidity, wind speed, seismic resilience, electromagnetic compliance, frequency offset, and other operational requirements.

**Keywords:** instrumentation: interferometers – methods: observational – telescopes

## 1 SKA PHASE SYNCHRONISATION

The Square Kilometre Array (SKA) (Dewdney et al., 2009; Carilli & Rawlings, 2004) is an international project to construct the world-leading radio radio telescope operating at frequencies between 50 MHz to 15.4 GHz. The first phase of the project (SKA1) will result in the construction of the initial 10% of the telescope's receiving capacity. In this phase, a low-frequency aperture array comprising over a hundred thousand individual dipole antenna elements will be constructed in Western Australia (SKA1-LOW), while a mid-frequency array of 197 parabolic-dish antennas, incorporating the 64 dishes of the MeerKAT telescope (Jonas, 2009), will be built in South Africa (SKA1-MID).

The SKA, being an interferometric array of individual antennas, requires coherent frequency reference signals at each antenna in the array to achieve phase coherence as required for performing interferometry and beamforming. In the case of the SKA1-MID, these reference signals will be generated centrally and transmitted to the antennas via fibre-optic links up to 175 km in length, with the longest links comprising mostly overhead cables (Grainge et al., 2017). Environmental perturbations add phase noise to the reference signals, degrading the array coherence, and thereby reducing the fidelity and dynamic range of the astronomical data (Cliche & Shillue, 2006).

Low frequency arrays, such as the Murchison Widefield Array (MWA) (Lonsdale et al., 2009; Tingay et al., 2013) and the Low-Frequency Array (LOFAR) (van Haarlem et al., 2013), or intermediate frequency arrays with a compact configuration, such as MeerKAT (Jonas, 2009; Julie & Abbott, 2017) and the Westerbork Synthesis Radio Telescope (Hogbom & Brouw, 1974), can attain sufficient phase coherence with a carefully designed passive frequency distribution system. In such systems, the frequency reference is simply transmitted to each antenna with no active attempt to control the fluctuations in phase. Arrays operating at higher frequencies and/or that are more extended in size, such as the Very Large Array (VLA) (Thompson et al., 1980) and the Australia Telescope Compact Array (ATNF, 2016; Hancock et al., 2011), often incorporate a round-trip phase measurement system. Here the system records changes in phase, and feeds forward a prediction of the future phase to the array's correlator. However, for both the mentioned arrays, the system is currently not in use.

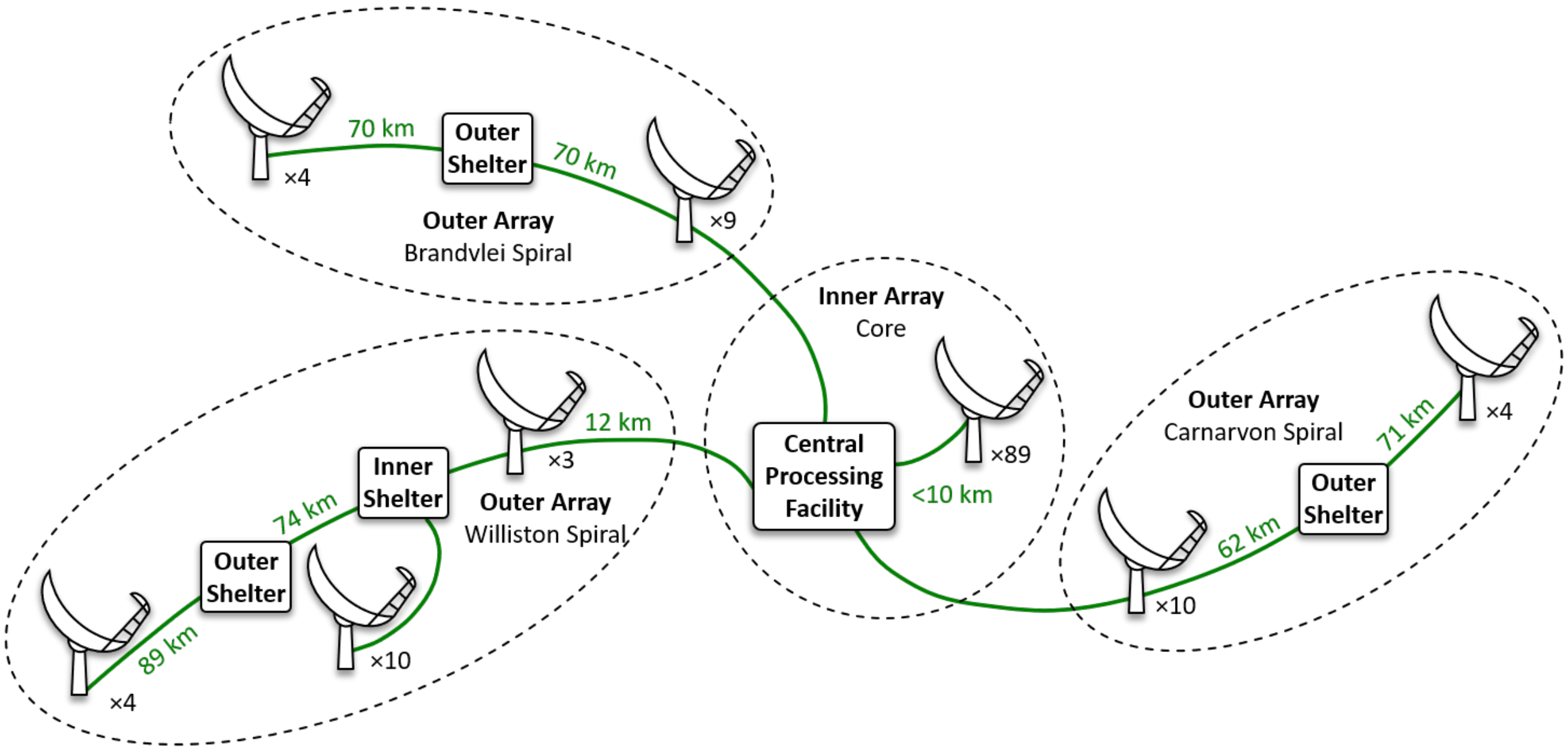

**Figure 1.** Layout of the planned SKA1-MID telescope showing the locations of the antennas in the core and in the three spiral arms.

Telescopes operating at high frequencies, such as the Atacama Large Millimetre Array (ALMA) (Wootten & Thompson, 2009; Shillue et al., 2004), or arrays that have a very large physical extent, such as the enhanced Multi-Element Radio Linked Interferometer Network (e-MERLIN) (McCool et al., 2008; Garrington et al., 2004), require actively-stabilised frequency transfer to suppress the fibre-optic link noise to maintain phase-coherence across the array. A similar scheme was also implemented using coaxial cable transmission lines for the compact, but high-frequency, Plateau de Bure Interferometer (Guilloteau et al., 1992). In addition, actively-stabilised frequency transfer has also been proposed for synchronising what are currently individually-referenced Very Long Baseline Interferometry (VLBI) antennas (He et al., 2018; Clivati et al., 2017; Krehlik et al., 2017; Dierikx et al., 2016)

Given the combination of long fibre links, particularly from the increased environmental exposure of overhead cables, and relatively high frequencies of the astronomical observations, the SKA1-MID will therefore employ actively-stabilised frequency transfer technologies to suppress the fibre-optic link noise to maintain phase-coherence across the array (Grainge et al., 2017). Given these unique challenges and many specific technical requirement of the SKA, a novel, bespoke phase synchronisation system had to be designed specifically for the SKA.

In this paper we describe the technical details and practical implementation of the SKA1-MID phase synchronisation system. Following a four-year development period, this system was formally selected for use on the SKA1-MID telescope in November 2017, and it has passed its Critical Design Review in October 2018. The system has been extensively tested using well-established metrology techniques while operating over installed fibre links (Schediwy et al., 2017) and during field trials on long-haul overhead optical fibre, and further verified using novel astronomical techniques with existing radio telescope arrays (Gozzard et al., 2017c).

## 2 STABILISED FREQUENCY TRANSFER

The SKA1-MID phase synchronisation system is based on the transmission of actively-stabilised microwave-frequency reference signals that are generated at the SKA's central processing facility (CPF), and then transmitted via separate fibre-optic links to each antenna. The star-shaped network topology of this phase synchronisation system, matches the fibre topology of the SKA's data network shown in Figure 1. This means that the same fibre-optic network that carries the astronomical data from the antennas to the CPF can also be used by the phase synchronisation system. The actively-stabilised frequency transfer technique at the core of SKA1-MID phase synchronisation system builds on the technology used in ALMA's distributed 'photonic local oscillator system' (Shillue et al., 2004), by incorporating advances made by the international frequency metrology community over the last decade (Foreman et al., 2007; Lopez et al., 2010b; Predehl et al., 2012), as well as several key innovations developed by the authors specifically for the SKA telescope (Schediwy et al., 2017; Gozzard

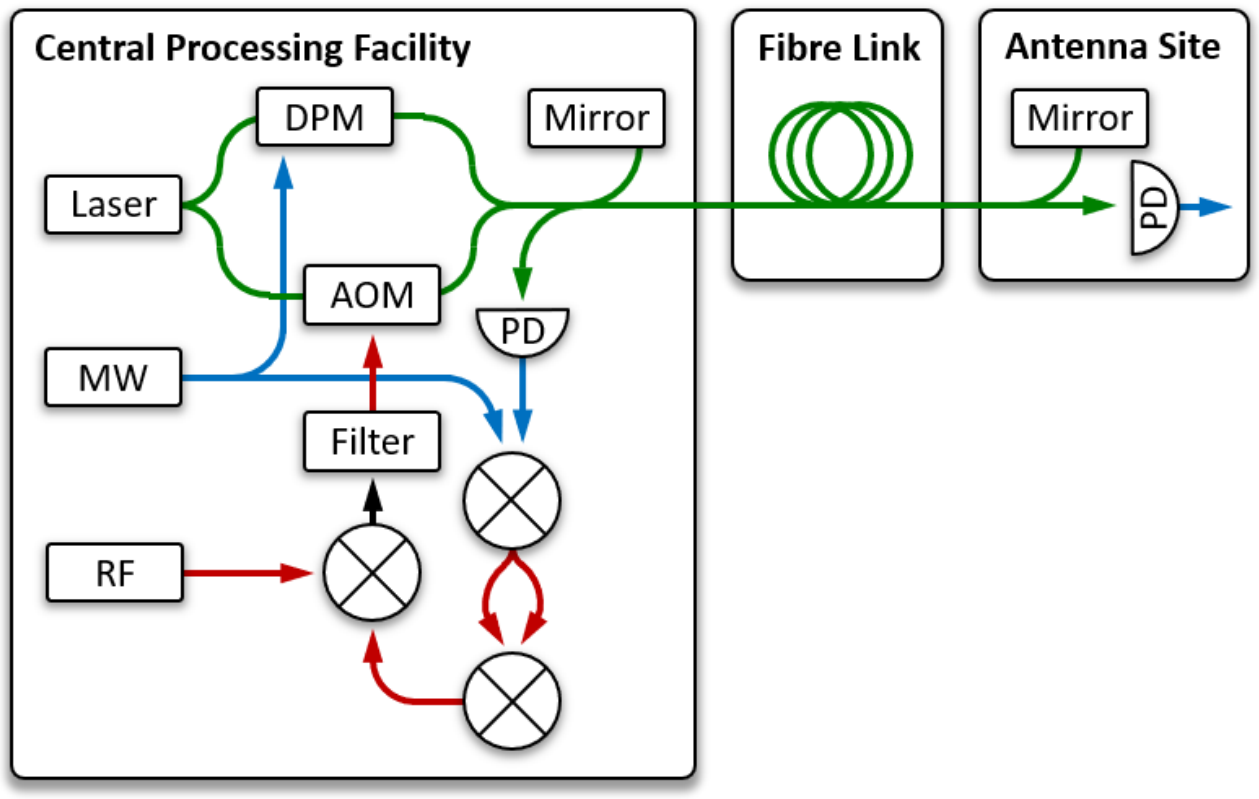

**Figure 2.** Simplified schematic diagram of the stabilised frequency transfer technique forming the core of the SKA1-MID phase synchronisation system. DPM, dual-parallel Mach-Zehnder modulator; AOM, acousto-optic modulator; PD, photodetector; MW, microwave-frequency source; RF, radio-frequency source.

et al., 2018b; Schediwy, 2012; Gozzard et al., 2018a).

The SKA1-MID phase synchronisation system transmits a microwave-frequency reference as a sinusoidal beat between two optical-frequency signals. As shown in Figure 2, the system is fully fiberised, and generates this microwave-frequency reference signal by physically splitting the optical signal from one laser into two arms of a fiberised Mach-Zehnder interferometer (MZI). In one arm of this interferometer, a microwave (MW) frequency shift is applied to the optical signal using a dual-parallel Mach-Zehnder modulator (DPM) whose input signals are biased in a manner to generate single-sideband suppressed-carrier (SSB-SC) modulation. This effectively produces a MW frequency shift of the optical signal. The other arm contains an acousto-optic modulator (AOM), which produces a radio-frequency shift on the optical signal.

When the two MZI arms are recombined, the result is a microwave-frequency beat between the two optical frequencies. This microwave-frequency beat signal is transmitted over the fibre-optic link to each of the antennas, where it is converted into the electronic domain by a photodetector. This electronic signal is then used as the frequency reference for the SKA1-MID receivers.

Active stabilisation of the transmitted signal is achieved through the use of an imbalanced Michelson interferometer (MI). At the antenna, a mirror reflects part of the optical signal back to the CPF forming the long arm of the MI. This reflected signal is then mixed with a copy of the transmitted optical signal originating from a short MI reference arm, and the resultant frequency modulation is extracted at the MI output using a photodetector. The electronic signal is then mixed with a microwave-frequency reference to produce two radio-frequency signals. These are mixed together, and their product is further mixed with a radio-frequency reference to produce an error signal that contains an imprint of the fluctuations of the fibre optic link. Appropriately filtering this signal and using it to steer the input frequency of the AOM, closes the servo loop and effectively supresses the phase noise acquired on the fibre link as well as the MZI. The full technical details of the stabilisation technique are given in (Schediwy et al., 2017); the remainder of this paper describes how this technique is used to build a phase synchronisation system designed specifically to meet the scientific needs and technical challenges of the SKA telescope.

## 3 SKA1-MID REQUIREMENTS

The SKA1-MID phase synchronisation system must meet a set of performance and operational requirements to ensure sufficiently high coherence between all the receptors to allow beam-forming and interferometry. The principal requirements of the SKA1-MID phase synchronisation system dictate that the distributed reference signals be stable enough, over all relevant timescales, to ensure a sufficiently-high coherence of astronomical data detected by the telescope array.

There are three specific phase stability performance requirements (Turner, 2015): The requirement for <1.9% coherence-loss within a maximum integration period of 1 second is dictated by the correlator integration time; while <1.9% coherence-loss over a 60-second integration period is determined by the period between in-beam calibrations. Finally, the requirement for <1 radian of phase drift over intervals up to 10 minutes is to ensure no wrap-around ambiguity in the phase solution between phase calibration observations. For the SKA1-MID phase synchronisation system, the dominant factor affecting the coherence loss is the choice of frequency of the transmitted reference signals. Therefore, the system is designed to transmit the highest MW frequency practicable given the constraints of its constituent key hardware elements.

The system must deliver the stabilised frequency reference signals to all 197 antenna sites across a range of expected operating conditions. The system must meet the above coherence loss requirements while subjected to temperatures between $-5^o$C and $+50^o$C within the dish pedestals and fibre link, and between $+18^o$C and $+26^o$C at the CPF and repeater shelters. In addition, the system must be able to operate with temperature gradients of $\pm 3^o$C every 10 minutes within the dish pedestals and on the fibre link; and in non-condensing relative humidity between 40% and 60% within the dish pedestals and repeater shelters. The system must also be able to operate over the maximum link distances of 175 km on overhead fibre, with wind speeds up to 40 km per hour. Finally, the system must be resilient to seismic shocks of 1 m/s$^2$.

Additional important requirements include the need for every individual antenna's reference signal to be transmitted with unique frequency-offsets that are integer multiples of exactly 10 kHz, to help the correlator to reject radio-frequency interference; and for the system to rigorously limit any self-generated radio-frequency transmission (this is especially critical for the equipment at the antenna site, as it will likely be located at the antenna indexer). There are also limits on power consumption and on available space. Space is severely constrained in both the CPF and at the antenna indexer. Operational requirements include aspects of system availability, operability, maintainability, and monitoring. Finally, the SKA1-MID phase synchronisation system must meet all these requirements for less than 1/10th of the cost per link compared to the equivalent ALMA system.

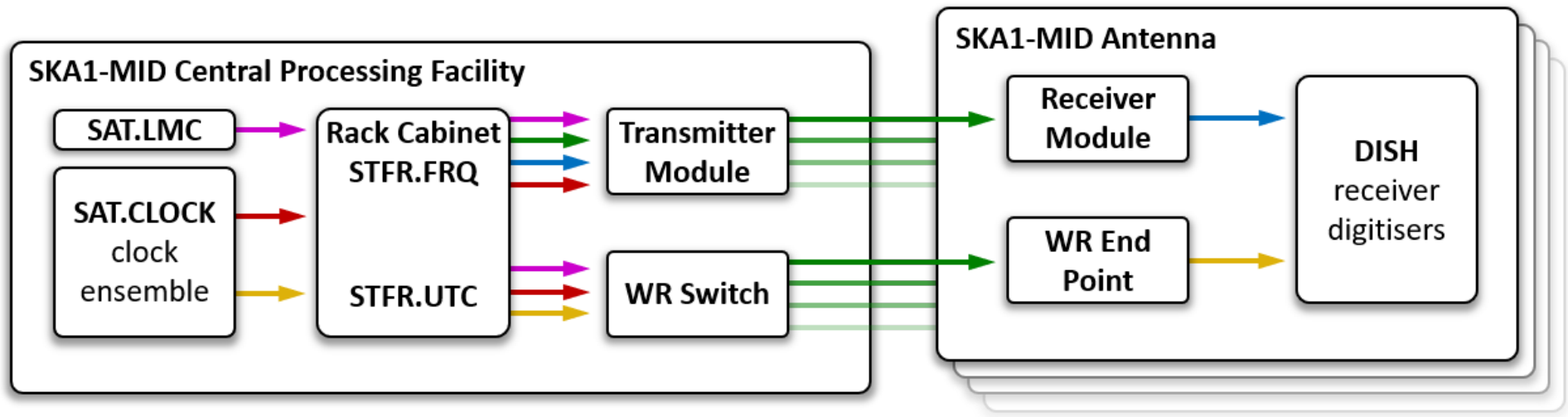

**Figure 3.** Schematic diagram of the SKA1-MID Synchronisation and Timing (SAT) network. Signal colour code: pink, data; red, radio frequencies; blue, microwave frequencies; green, optical frequencies/optical fibre; yellow, timing signals.

## 4 PRACTICAL IMPLEMENTATION

The SKA1-MID phase synchronisation system is an element within the SKA Signal and Data Transport (SADT) consortium's Synchronisation and Timing (SAT) network. The full SADT element name of the SKA1-MID phase synchronisation system is SADT.SAT.STFR.FRQ, where STFR is an acronym of 'Station Time and Frequency Reference' and FRQ is an abbreviation of 'Frequency'. The other elements within the SAT network are SAT.CLOCK, the SKA's hydrogen maser clock ensemble and timescale; SAT.UTC, the system for disseminating absolute time (which utilises 'White Rabbit' (WR) technology (Dierikx et al., 2016; Boven, 2017); and SAT.LMC, the local monitor and control system for the SAT network. The output of SADT.SAT.STFR.FRQ is passed to the SKA1-MID receivers, which are the remit of the DISH consortium. A simplified schematic of the SADT.SAT network architecture, showing the inter-relationships between the various SADT.SAT network elements and DISH, is given in Figure 3.

The SKA1-MID phase synchronisation system comprises a combination of ten hot-swappable modules. The individual modules for transmitting and receiving the frequency reference signals (the **Transmitter Module** and **Receiver Module** respectively), as well as the required mounting and racking hardware (**Rack Cabinet**) are shown in Figure 3. As shown in Figure 4, the **Rack Cabinet** incorporates the transmission laser (**Optical Source**), the microwave-frequency source (**Frequency Synthesiser**), the module for encoding the microwave-frequency signal into the optical domain (**Microwave Shift**), the reference radio-frequency source (**Signal Generator**), and some passive signal distribution modules (**Rack Distribution**). In addition, the **Rack Cabinet** holds multiple **Sub Rack** modules, each of which holds 16 **Transmitter Modules**. The simplified schematics for the **Transmitter Module** and **Receiver Module** are shown in Figure 5.

The **Rack Cabinet** takes input from SAT.CLOCKS and SAT.LMC, and produces a series of outputs that are passed to the **Transmitter Module**. The **Transmitter Module** transmits the reference signal across the fibre link to the **Receiver Module**. The **Transmitter Module** contains the servo-loop electronics and all monitor and control hardware. The **Receiver Module** includes a clean-up oscillator in a phase-locked loop, with the output then passed onto the SKA1-MID antenna receivers. Note, for the longest 14 of the total 197 SKA1-MID fibre links, optical amplification is required to boost the frequency reference signal strength; the **Optical Amplifiers** (not shown in the figures) are therefore located in various repeater shelters and antennas pedestals (see Schediwy & Gozzard (2018) for details).

Four of these modules are available commercially-off-the-shelf (COTS), and the remaining six modules are bespoke assemblies made up of mainly COTS components. During the SKA Pre-Construction stage, the SKA project intends to keep the telescope designs 'vendor agnostic', and therefore no specific products are mentioned in the following section. Suitable components can be sourced from a range of vendors.

## 5 IMPLEMENTATION DETAILS

The specific implementation of the SKA1-MID phase synchronisation system is as follows: As per Figure 4, an optical signal with a frequency of 193 THz, generated by the **Optical Source**, is fed into a fiberised MZI. A microwave-frequency reference signal of 7.96 GHz generated by the **Frequency Synthesiser** is applied to the DPM incorporated within the **Microwave Shift** module, which is tuned to produce a static microwave-frequency shift of the optical signal, producing a new optical signal with a frequency of 193 THz − 7.96 GHz. As shown in Figure 4, these two optical-frequency signals are split 256 ways to provide the required inputs for all of the system's 197 active **Transmitter Modules** (the system also maintains several 'hot' spares).

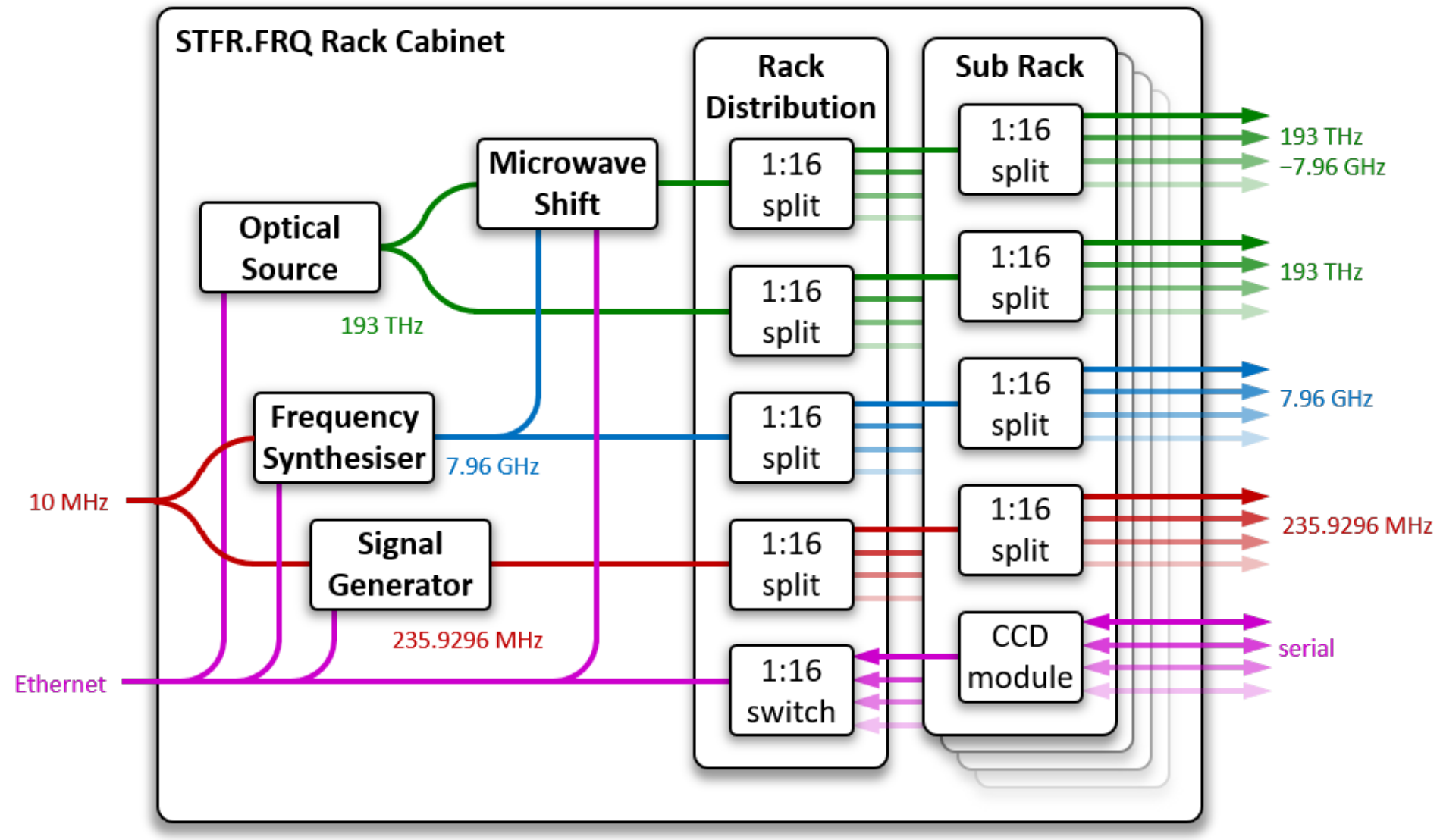

**Figure 4.** Schematic diagram of the SKA1-MID STFR.FRQ **Rack Cabinet**.

In each **Transmitter Module**, the 'servo' acousto-optic modulator (AOM), adds +40 MHz to the 193 THz optical signal. This signal is then recombined with the DPM-shifted optical signal at the output of the MZI, and the signals pass an optical isolator (Iso.) to prevent any reflections from returning to the laser. The signals then enter an unbalanced MI, where the fibre link is the long interferometer arm, and the short interferometer arm acts as the physical reference for the optical phase. The MI reference arm contains a local 'anti-reflection' AOM, which is used to apply a static frequency offset of −85 MHz that allows the servo electronics to reject unwanted reflections from the link. Therefore, the optical reference signals at the photodetector have frequencies of 193 THz − 8.13 GHz and 193 THz − 130 MHz.

Two optical signals are transmitted across the fibre link; these have frequencies of 193 THz − 7.96 GHz and 193 THz + 40 MHz. As the signals are transmitted over the fibre link, they pick up phase noise resulting from optical path length changes occuring in the link due to environmental fluctuations. At the **Receiver Module**, a remote 'anti-reflection' AOM applies a static frequency of +75 MHz, produced by a simple crystal oscillator (XO), to the two optical signals. The signals are then split, with one set of signals being received at a photodetector (PD), where the beat note results in an 8.00 GHz remote electronic signal. The second set set of optical signals reflects off a Faraday mirror (FM) back across the fibre link to the **Transmitter Module**. Each optical signal receives an additional shift of +75 MHz when passing through the remote AOM a second time. The optical signals returning to the **Transmitter Module** now have frequencies of 193 THz − 7.81 GHz and 193 THz + 190 MHz.

When the two reference frequency signals are mixed with the two reflected optical frequency signals, this results in six primary electronic mixing products. Of those six, the two microwave-frequency mixing products with frequencies of 8.32 GHz and 7.68 GHz, contain the combination of phase noise information required by the system. The six mixing products are then mixed with the microwave-frequency reference signal at 7.96 GHz. The two signals carrying the phase noise information produce radio-frequency signals at 360 MHz and 280 MHz. An electronic MZI is used to separate the signals into two physical paths, with each path containing a band-pass filter (BPF) that is centred on one of the above frequencies. These filters reject the unwanted mixing products and any frequency inter-modulations. The two radio-frequency signals are then mixed and filtered to produce a signal with a frequency of 80 MHz. Finally, the 80 MHz signal is mixed with the local oscillator (LO) signal at 80 MHz, to produce the DC servo error signal. The LO signal is generated by a direct digital synthesiser (DDS) with a clock of 235.9296 MHz. This clock frequency is produced by the **Signal Generator**, which is referenced by a 10 MHz signal from the SKA's

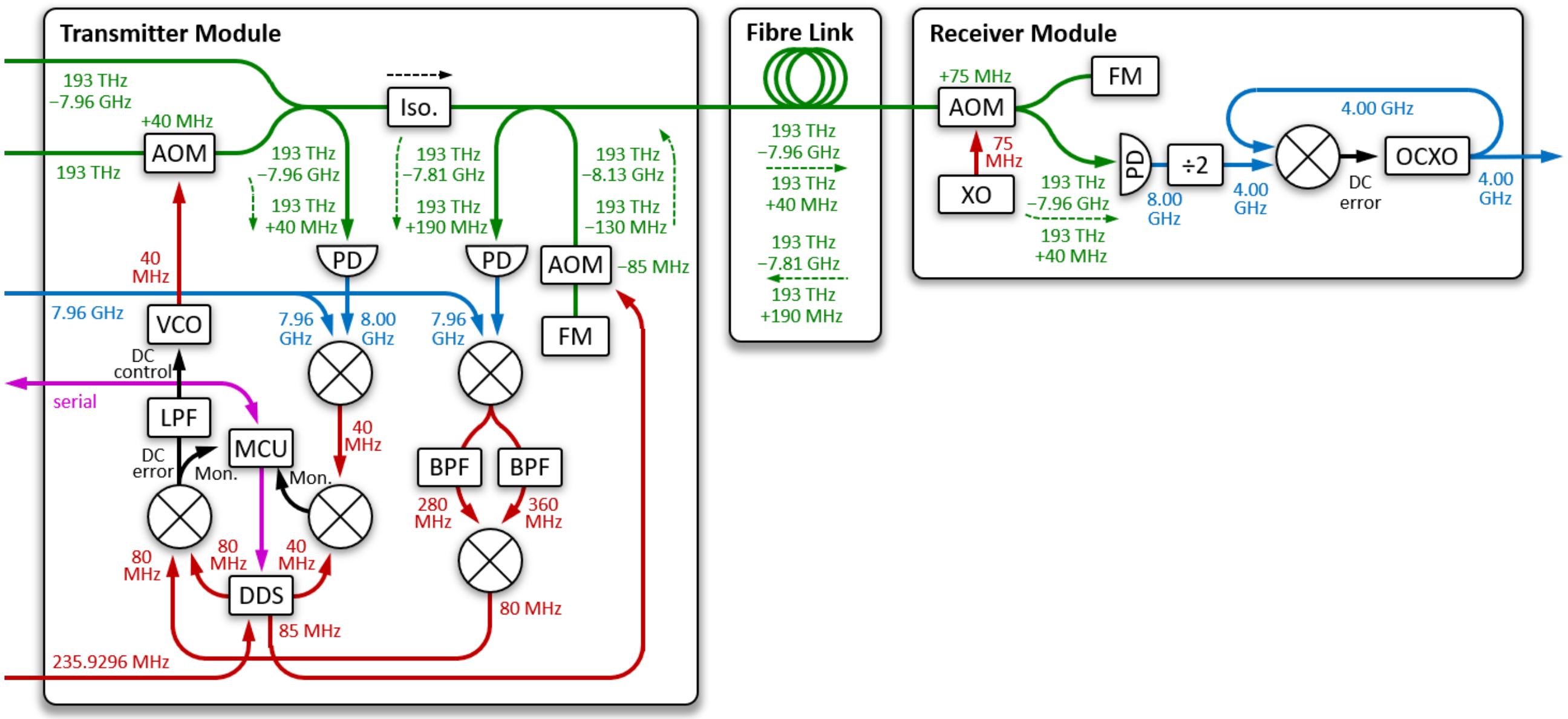

**Figure 5.** Schematic diagram of the SKA1-MID **Transmitter Module** and **Receiver Module**. AOM, acousto-optic modulator; PD, photodetector; Iso., optical isolator; FM, Faraday mirror; VCO, voltage-controlled oscillator; LPF, low-pass filter; MCU, microcontroller unit; BPF, band-pass filter; DDS, direct digital synthesiser; XO, crystal oscillator; ÷2, frequency divider; OCXO, oven-controlled crystal oscillator.

SAT.CLOCK hydrogen-maser clock ensemble.

The DC error signal is low-pass filtered to reject higher-frequency mixing products and then fed into a voltage-controlled oscillator (VCO). The VCO has a nominal centre frequency of 40 MHz. The VCO signal is used to drive the servo AOM, with variations in the DC error signal mapping directly to changes in AOM frequency. The servo loop attempts to drive the error signal to zero volts. As demonstrated in Schediwy et al. (2017), this has the effect of suppressing any phase noise added by the fibre link, as well as the MZI, that is stationary at timescales slower than the light round trip time. The DC error signal is also recorded using a microcontroller unit (MCU), and then relayed to SAT.LMC.

At the **Receiver Module**, the 8.00 GHz signal is divided by two (÷2) to produce 4.00 GHz. This signal is mixed with a signal of the same frequency produced by an oven-controlled crystal oscillator (OCXO). The resultant low-pass filtered DC signal is fed into the control port of the OCXO. This creates a phase-locked loop (PLL) where the 4.00 GHz output signal tracks the transmitted reference signal at frequencies below the PLL cut-off frequency, while the short timescale phase noise is dominated by the intrinsic stability of the OCXO. The optimal combined phase noise is expected when the PLL cut-off frequency is set below the frequency-equivalent light round trip time of the fibre link. As the longest SKA1-MID fibre links are 175 km, the light round trip time will be around 1.7 ms, resulting in ideal maximum loop bandwidth of around 570 Hz. However, practically, the optimal loop bandwidth may be between 10 and 100 Hz. The 4.00 GHz OCXO output signal is passed to the SKA's DISH receiver digitisers.

The SKA1-MID phase synchronisation system also includes an out-of-loop monitoring system. As shown in Figure 5, the transmitted optical signals are detected with a second 'monitor PD' incorporated into the **Transmitter Module**. The resultant 8.00 GHz electronic signal is mixed with the 7.96 GHz signal generated by the **Frequency Synthesiser**, to produce a 40 MHz mixing product. This is further mixed with a 40 MHz reference signal from the DDS to produce a DC monitor signal. Again, the MCU is used to record this signal and passes it to SAT.LMC.

## 6 SYSTEM PERFORMANCE

The functional performance of the SKA1-MID phase synchronisation system was experimentally verified using the following method. An independent out-of-loop measurement of the system's transfer stability was able to be made by locating the and **Receiver Module** in the same laboratory. The frequency reference signal was transmitted from the **Transmitter Module** to the **Receiver Module** over up to 166 km of optical fibre networks (two 83 km fibre loops, managed by AARNet) installed in a metropolitan area. The two-way light round trip time for the signal transmission was 1.6 ms, thereby limiting the servo bandwidth to around 600 Hz. The total optical loss of the link was measured to be 47 dB.

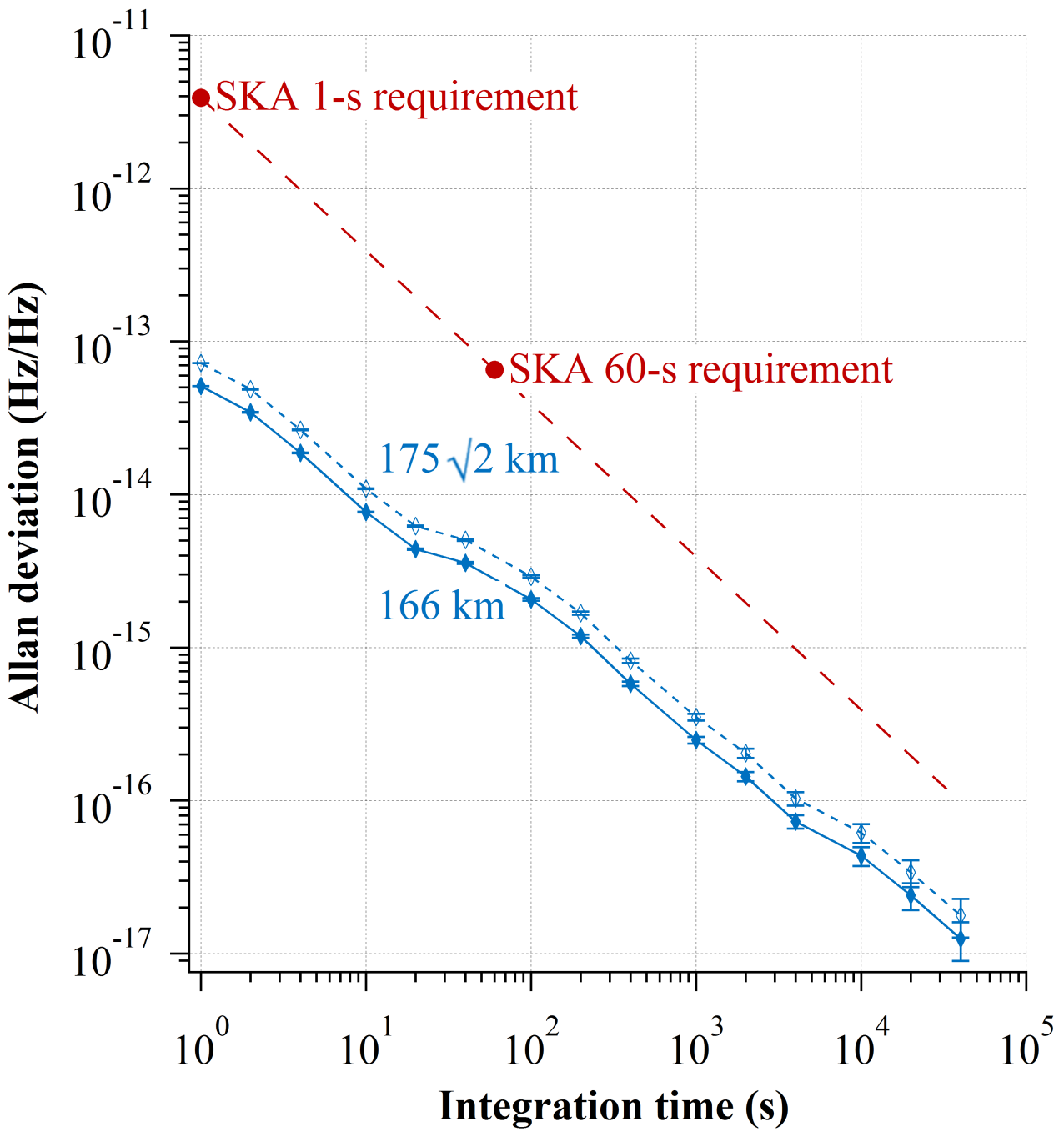

**Figure 6.** Allan deviation as a function of integration time of the SKA1-MID phase synchronisation system as measured over a 166 km metropolitan fibre link (solid blue line, filled diamonds), and estimated for two independent antennas with the maximum fibre distance of 175 km (dotted blue line, open diamonds). Also shown are the two specified coherence requirements converted to an Allan deviation (filled red circles).

Two bi-directional optical amplifiers manufactured by IDIL Fibres Optiques with a total gain of 34 dB were used to partially make up for the signal attenuation on the fibre link. The first amplifier was located after the initial 83 km fibre loop, and the second amplifier just prior to the **Receiver Module**. The 8.00 GHz remote electronic signal was mixed with the 7.96 GHz signal generated by the **Frequency Synthesiser** to produce a 40 MHz signal.

### 6.1 Coherence Loss Requirement

For the evaluation of the coherence loss, the 40 MHz signal was probed using a Microsemi 5125A Test Set. The raw Allan deviation data produced by the Test Set were divided by the mixing ratio of 200 (8.00 GHz/40 MHz) to obtain the appropriate fractional frequency stability. Figure 6 shows a plot of these computed Allan deviation values.

The most stringent use case of the SKA1-MID phase synchronisation system is synchronisation of two antennas each located at the end of separate 175 km links. To predict the coherence loss performance for this use case, the Allan deviation values for the 166 km measurements were first extrapolated to 175 km using the process described in Williams et al. (2008). The measured Allan deviation values are multiplied by $(L2/L1)^{3/2}$, where L1 is the length of the measured link (166 km) and L2 is the extrapolated length (175 km). Secondly, to predict the stability between two independent systems, the extrapolated Allan deviation values were then multiplied by a factor of $\sqrt{2}$.

The Allan deviation measurement has a slope with a gradient close to $\tau^{-1}$. This indicates that the dominant noise process affecting the stabilised frequency reference is white phase noise. An estimate of the coherence loss can therefore be calculated using the process described by Thompson et al. (2017, p. 435) and Rogers & Moran (1981). With this, a permissible Allan deviation upper limit value of $3.9\times10^{-12}/\tau$ was calculated given the requirement of 1.9% coherence loss at the maximum observing frequency.

If the same coherence calculations are applied to the extrapolated Allan deviation values, the coherence losses are determined to be $7.41\times10^{-6}$ and $7.96\times10^{-6}$ at one second and 60 seconds of integration time respectively. This results in an excess of the coherence loss requirement by a factor of 2560 at one second of integration, and a factor of 239 at 60 seconds.

### 6.2 Phase Drift Performance

The phase drift performance was measured by mixing down the aforementioned 40 MHz out-of-loop measurement signal to DC using a stable 40 MHz LO. This 40 MHz LO was derived by frequency dividing the existing 80 MHz LO by a factor of two. Both the RF and LO inputs of the mixer were band-pass filtered and the input powers were tuned to drive the mixer at its optimal level.

Temporarily using an independent RF input set at identical power level, but detuned from the LO by 50 kHz, revealed that the discriminator curve was approximated by an ideal sinusoidal shape, with a peak-to-peak range of 274 mV. This produced a phase-to-voltage discriminator with a gradient of 137 mV/rad, which is 99% linear within a range of ±53.4 mV (±0.39 rad).

With the measurement signal re-connected, the phase of the LO signal was adjusted using a variable delay line to centre the mixer DC output as close as possible to 0 V. This was necessary to ensure that any deviations would remain within the linear output range of the mixer.

With this initialisation process completed, the resultant DC signal was logged using an Agilent 34410A digital multimeter. The root mean square of the measured voltage data is 58 $\mu$V, so well within the linear range of the mixer. The recorded time-series values were converted to phase deviations using the discriminator value calculated above, and the resultant absolute phase time-series was then used to calculate the magnitude of phase drift for each 10 minute period within the data set. The magnitude of phase drifts were tallied and the results are presented as a histogram in Figure 7.

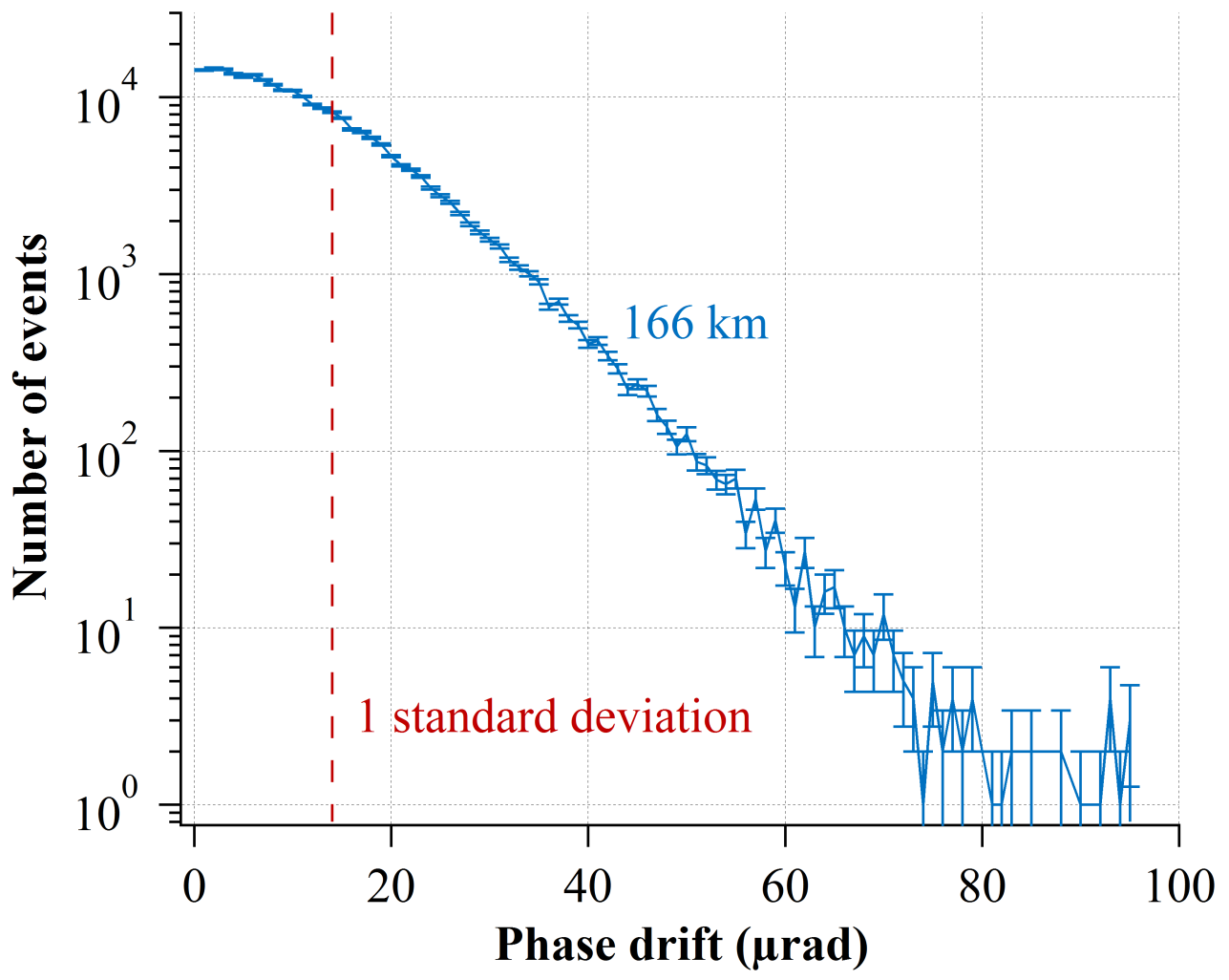

**Figure 7.** Histogram of 10-minute phase drift magnitude of the SKA1-MID phase synchronisation system shown with 1 $\mu$rad bins, as measured over a 166 km metropolitan fibre link (solid blue line). Also shown is the 1-standard deviation cut-off value of 14 $\mu$rad (dashed red line); the requirement is for <1 rad.

The 1-standard deviation value of 10 minute phase drift of the SKA1-MID system deployed over a 166 km urban conduit optical fibre link is 14 $\mu$rad. There were no 10 minute phase drift events greater than 100 $\mu$rad during the entire 15 hour test period. The phase synchronisation system exceeds the 10 minute phase drift requirement by almost five orders of magnitude.

### 6.3 Normal Operating Conditions

In addition to the above mentioned baseline measurements, the transfer stability performance of the system was tested repeatedly over a wide range of normal operating conditions. The 1 second and 60 second coherence loss, as well as the 10 minute phase drift values were determined to exceed SKA performance requirements over all of the SKA specified normal operating conditions outlined below. Further details of the particular experimental methods of each test and the associated specific performance values are outlined in the SKA1-MID Detailed Design Report (Schediwy & Gozzard, 2018). These normal operating conditions are stipulated according to the location of the equipment, with the defined locations being the CPF, the shelters (see Figure 1), and the pedestal at the antenna site. For a breakdown of the location of each equipment see Figure 3 and Figure 4.

The specified operating condition for temperature range are $-5^oC$ to $+50^oC$ at the antenna site and for the fibre link, and $+18^oC$ to $+26^oC$ for the CPF and shelters. The CPF equipment was tested in an environmental test chamber over a range of $-8^oC$ to $+51^oC$; the shelter equipment between $-8^oC$ to $+57^oC$, and the equipment in the CPF between $+16^oC$ to $+30^oC$. During a two-week field deployment of the system on long-haul overhead fibre at the South African SKA site (Gozzard et al., 2017b), the ambient temperature of the 186 km fibre link unfortunately ranged only between $0.9^oC$ to $14.8^oC$ (Gozzard et al., 2017a).

The maximum temperature gradient is specified only for the antenna pedestal and fibre link. It is $\pm3^oC$ over a 10 minute period. Again an environmental test chamber was used to for equipment at the antenna pedestal, achieving maximum temperature gradients of at least $+3^oC$ and $-13^oC$ over a 10 minute period. The maximum temperature gradient experienced by the 186 km fibre link during the field deployment was only $\pm1.2^oC$ over 10 minutes.

The requirement for non-condensing relative humidity is between 40% and 60% at all locations. Test were conducted with all equipment exposed to a relative humidity range between 9% and 99%. The fibre link experienced a relative humidity between 21% and 97% during the field deployment. The final operating conditions specify that the system has to operate on overhead fibre with wind speeds up to 40 km per hour. During the field deployment, the system was operated and tested at wind speeds up to 50 km per hour.

In addition to the above normal operating conditions, a few other system properties were experimentally investigated. It is required that the system can survive seismic shocks up to 1 $m/s^2$. This was verified by loading the equipment in a vehicle and subjecting the equipment to accelerations in all three axes at a variety of frequencies, including shock loads, up to 10 $m/s^2$.

Finally, the system was shown to exceed both the broadband and narrowband electromagnetic compliance requirements over the entire specified frequency range. The CFP equipment exceed specifications by a minimum of 27 dB for broadband, and 14 dB for narrowband emission, and the antenna hardware by a minimum 50 dB for broadband, and 37 dB for narrowband emission.

### 6.4 Additional Design Elements

The SKA1-MID telescope will employ a sample clock frequency offset scheme in order to assist with the mitigation of strong out-of-band radio-frequency interference and digitiser artefacts. The SKA1-MID phase synchronisation system will need to deliver the nominal microwave-frequency reference signal to each antenna site with an offset determined by a unique integer multiple of 10 kHz. The system achieves this by simply commanding the DDS to apply the appropriate offset to the servo LO in each **Transmitter Module**. This then changes the control signal applied to the servo AOM, which in combination with the fixed microwave frequency applied in the **Microwave Shift** module, applies the appropriate offset to the frequency reference sent to the antennas.

The equipment in the CPF fits within two rack cabinets. The **Optical Source**, **Microwave Shift**, **Frequency Source**, **Signal Generator**, and **Rack Distribution** require 15 rack units (U) in total. In addition, the CPF equipment includes 14, 3 U-high **Sub Racks**, each of which house 16 Transmitter Modules; 100 mm high, 580 mm deep Eurocard printed circuit boards (PCB). A few other rack units are dedicated to cooling, cable managements, network switches, and rack power supplies. The commercial bidirectional amplifiers in the repeater huts are 1 U. The optical and electronic components in the **Receiver Module** at the antenna site have a volume of less than 1.5 litres. The equipment in the CPF requires 2.3 kW of electrical power. Each of the 14 bidirectional amplifiers in the repeater shelters or antenna pedestal require no more than 50 W each. The **Receiver Module** is evaluated to require 5 W.

The MCU on each **Transmitter Module** is used to monitor the out-of-loop monitoring system and DC error signal, reporting the values to SAT.LMC at a cadence of once per second. The MCU is also used to command the DDS to adjust the servo LO frequency to the appropriate value required by the frequency offset scheme. Each type of module is designed to be hot swappable and bespoke modules are constructed using mostly commercial-off-the-shelf components to help maximise maintainability, operability, and availability. All this is achieved with hardware costs of less than €10,000 per link.

## 7 DISCUSSION

The SKA1-MID phase synchronisation system was designed specifically to meet the scientific needs and technical challenges of the SKA telescope. The system is able to optimise key design parameters, while exceeding the performance needs of the SKA. The key innovation of SKA1-MID phase synchronisation system was finding a way to use AOMs as servo-loop actuators for modulated microwave-frequency transfer (Gozzard et al., 2018b). This is important because AOMs have large servo bandwidths as well an infinite phase actuation range. This means that practical systems can be made extremely robust, as their servo loop will never require an integrator reset; this is a common problem encountered by most other techniques. Previously, AOMs had only been used as actuators for optical frequency transfer (Schediwy et al., 2013; Ma et al., 1994).

For comparison, all 'standard' modulated frequency transfer techniques (Lopez et al., 2010b), including ALMA's (Shillue et al., 2012), require group-delay actuation to compensate the physical length changes of the fibre link. For practical deployments, this usually involves a combination of fibre stretcher (medium actuation speed and very limited actuation range) in series with at least one thermal spool (slow actuation speed and physically bulky). For the number of fibres, link distances, and servo loop actuation ranges encountered by the SKA, especially for SKA1-MID overhead fibre, which required actuation ranges hundreds of times greater than comparable-length underground fibre (Gozzard et al., 2017b), standard modulated frequency transfer is totally impractical. A novel solution to this is to conduct the group delay in the electronic domain (Krehlik et al., 2016); however, so far this has reportedly only been achieved using independent delay lines for the outgoing and return signals, thereby limiting the transfer performance below that of other leading techniques. Modulated 'phase conjugation' techniques (Primas et al., 1988) have also been successfully deployed over longer distances (He et al., 2013, 2018) than standard modulated frequency transfer, and a particular implementation of this technique was selected as the phase synchronisation system for SKA1-LOW (Wang et al., 2015).

In addition, the AOMs generate frequency shifts to mitigate against unwanted reflections that are inevitably present on real-world links. This reflection mitigation strategy cannot intrinsically be implemented with modulated frequency transfer techniques, as the carrier and sidebands would be shifted by the same frequency. Therefore, these modulated transfer techniques require the returned signal to be rebroadcast at either a different modulation frequency, optical wavelength, or on a different fibre core, to avoid frequency overlap from unwanted reflections on the link. These reflection mitigating methods thereby reduce the symmetry between outgoing and returning signals, leading to degraded stabilisation performance, and can bring about additional complications, including those resulting from optical polarization mismatch or from chromatic dispersion, which in turn requires further system complexity to overcome. However, reflection mitigation is essential for the SKA, as there is no way to guarantee that all links will remain completely free of reflections over the lifetime of the telescope.

The SKA1-MID system also benefits from being able to simultaneously suppress phase noise acquired on the fibre link and in the optical MZI contained within the CPF equipment (Schediwy et al., 2017). This makes the equipment extremely robust to environmental perturbations, such as the vibration generated by the large number of cooling fans within the CPF. In contrast, the transmitter rack of the ALMA photonic system had to be specifically engineered to not use any cooling fans, and physically isolated from other telescope equipment.

Another innovation of the SKA1-MID phase synchronisation system is the use of a DPM in parallel with an AOM to generate the microwave-frequency reference signal as a difference between two optical signals. Given the lower operating frequency of the SKA compared to ALMA (Shillue et al., 2004), the two optical signals can be generated using a single laser in combination with the DPM. This avoids the differential phase noise and complexity associated with offset-locking a set of independent master and slave lasers as per the ALMA system. Another method of using a DPM to generate two (widely-spaced) optical frequencies was proposed (Kiuchi et al., 2007), but ultimately not adopted, as an alternative photonic system for ALMA (Kiuchi & Kawanishi, 2009). However, this technique still relied on group-delay actuation and would therefore be incompatible with the technique described in this paper.

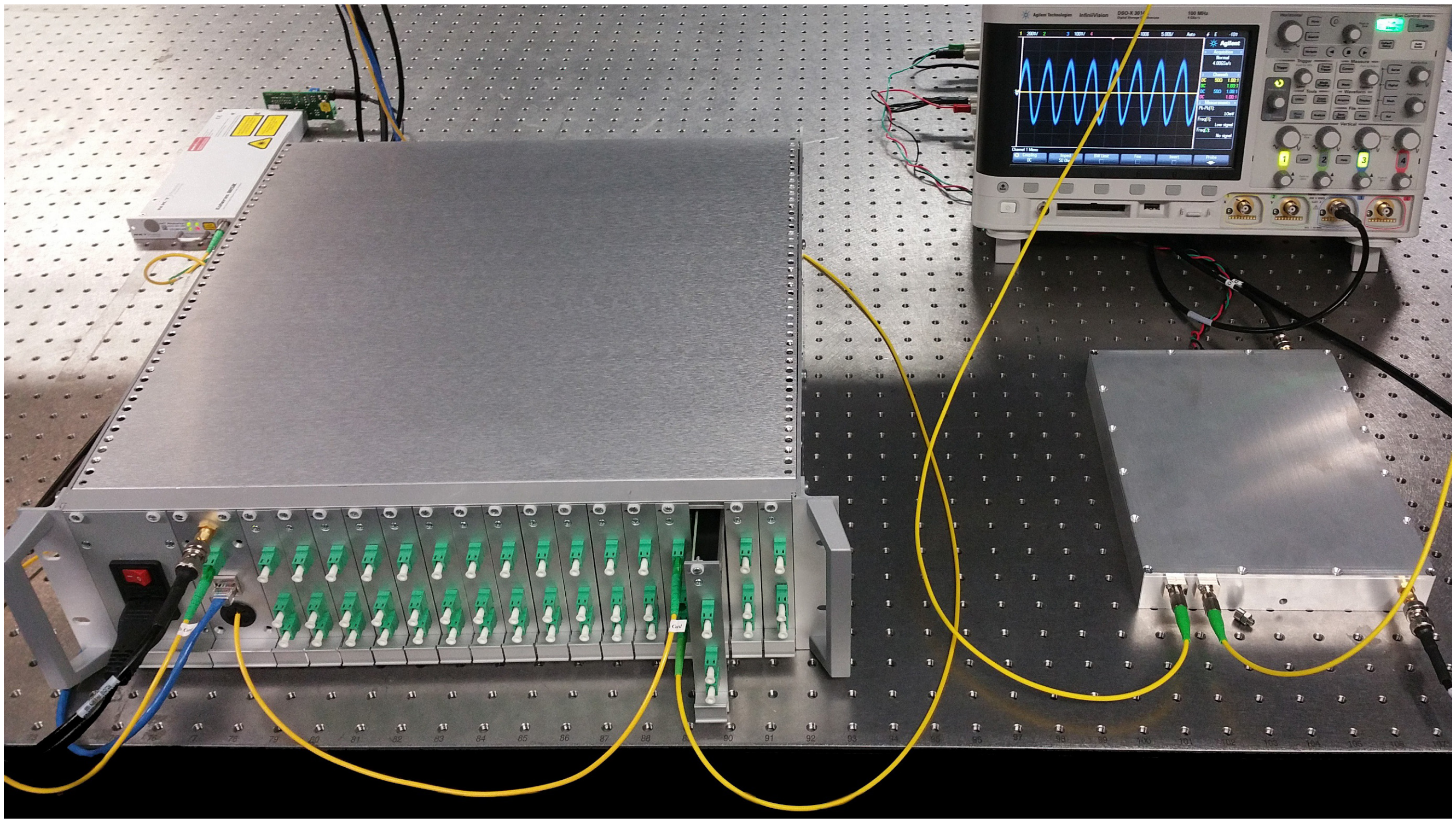

**Figure 8.** Photo of mass-manufacture prototypes of the SKA1-MID phase synchronising system. A **Transmitter Module** is shown protruding from the **Sub Rack** on the left, and the **Receiver Module** is shown on the right.

Similarly, using a single radio-frequency source (the **Frequency Generator**) to generate the reference signals for all independent **Transmitter Modules** avoids the phase drift that can occur even between two high-quality synthesisers (Gozzard et al., 2017c). This is because for a synthesiser to maintain a constant output frequency relative to an input reference, it must change the phase of that output frequency signal as its internal temperature changes in response to variations in ambient temperature.

Another benefit of a technique that transmits only two optical signals is the avoidance of signal fading normally encountered with ‘standard’ and ‘phase conjugation’ techniques that use dual-sideband modulation (Foreman et al., 2007). In these cases the chromatic dispersion of the fibre results in a differential evolution of phase of the two sideband-carrier products, resulting in periodic distances where the signal amplitude fades to zero amplitude. The SKA1-MID system intrinsically delivers the reference signal with maximum amplitude for any length of fibre.

Similarly, the use of Faraday mirrors as reflectors for the remote site and MI reference arm (as opposed to using standard mirrors or circulators), ensures that the signals are always aligned in polarisation when combined on the local photodetector (Lopez et al., 2010a). This avoids complexity and cost of a polarisation scrambler or polarisation alignment servo system (Lopez et al., 2010b). Optical amplifiers are used for simply boosting signal strength of the optical signals without having to have the complexity of an optical-electronic-optical conversion typical in a full regeneration system (Gao et al., 2015).

These advantages ensure that SKA1-MID phase synchronisation system can easily meet the functional performance requirements in practical realisation of the system while achieving maximum robustness, minimum cost, simple installation, and easy maintainability.

## 8 CONCLUSIONS

In this paper we have described the phase synchronisation system of the SKA1-MID telescope. Extensive test-ing has shown that the worst-case coherence loss for the SKA1-MID phase synchronisation system is $7.41\times10^{-6}$ at one second integration time, and $7.96\times10^{-5}$ at 60 seconds integration time. These values exceed the 1.9% coherence loss requirement by a factor 2560 and 239 respectively. Similarly, it was found that the system typically drifts in phase over a 10 minute period by no more than 14 $\mu$rad, exceeding the phase drift requirement by almost five orders of magnitude.

The SKA1-MID phase synchronisation system can operate within specification over all required operating conditions, including temperature range, temperature gradient, and relative humidity. In addition, the system was operated on overhead fibre links longer than the maximum SKA1-MID distance, and at wind speeds up to 50 km per hour. Finally, the tests have also demonstrated seismic resilience and electromagnetic compliance.

With the four-year development period completed and the SKA1-MID phase synchronisation system having passed its Critical Design Review, work has now progressed to completing the mass-manufacture prototypes in readiness for SKA Construction (see Figure 8).

## 9 ACKNOWLEDGEMENTS

This paper describes work carried out for the SKA Signal and Data Transport (SaDT) consortium as part of the Square Kilometre Array (SKA) project. The authors wish to thank Bassem Alachkar for useful discussions on interferometer coherence, and Thea Pulbrook for contributing to follow-up work stemming from the work reported here. Thanks also to AARNet for the provision of light-level access to their fibre network infrastructure.

The SKA project is an international effort to build the world's largest radio telescope, led by the SKA Organisation with the support of 10 member countries. This work was supported by funds from the University of Manchester and the University of Western Australia.